\documentclass{aa}
\usepackage[varg]{txfonts}
\usepackage{lineno} 
\usepackage{amssymb,amsmath}
\usepackage{graphicx}
\usepackage{xcolor,natbib}
\usepackage{multirow}
\usepackage[T1]{fontenc}
\usepackage{ae,aecompl}
\usepackage{newtxtext, newtxmath}
\usepackage{float}
\usepackage{subfigure}
\usepackage{longtable}
\usepackage{enumitem}
\usepackage{longtable,listings}
\usepackage[flushleft]{threeparttable}
\usepackage{parcolumns}
\usepackage{hyperref}
\usepackage{CJK}
\usepackage{upgreek}
\usepackage{soul}

\def\oiii{[O~{\sc iii}]}

\begin{document}

\title{Aperture-dependent \oiii~profile variability: Clues to the nature of double-peaked narrow emission lines}

\titlerunning{aperture effect}

\author{Qi Zheng \inst{\ref{inst}}
\and Xingqian Chen\inst{\ref{inst1}}
\and Zhegeng Chen\inst{\ref{inst1}}     
\and Xue-Guang Zhang\inst{\ref{inst1}}
\and Qirong Yuan\inst{\ref{inst}}}

\institute{School of Physics and Technology, Nanjing Normal University, No. 1,      Wenyuan Road, Nanjing, 210023, P. R. China\label{inst} \\
\and Guangxi Key Laboratory for Relativistic Astrophysics, School of Physical Science and Technology, Guangxi University, Nanning, 530004, P. R. China\label{inst1} \\
\email{xgzhang@gxu.edu.cn; yuanqirong@njnu.edu.cn}}
\date{Received 10/07/2026 / Accepted 27/10/2026}

\abstract
{
In this work we propose a simple and practical aperture-dependent method for investigating the physical origin of double-peaked narrow emission lines (DPNELs). Spectroscopic observations obtained from different apertures probe different spatial regions and therefore different gas kinematics within the narrow line region (NLR). Consequently, changes in the profiles of DPNELs observed with different apertures can provide important constraints on their physical origins.
We applied this method to SDSS J094144.83+575123.6 (SDSS J0941) by comparing spectra obtained from the SDSS, DESI, and LAMOST surveys. Significant differences are found in the double-peaked profiles of the different spectra, demonstrating that aperture effects can strongly influence the integrated DPNEL profiles. 
The dual active galactic nucleus (AGN) model, where the two nuclei lie in a plane perpendicular to the line of sight, can qualitatively explain the asymmetric sampling of the two NLRs, but it cannot explain the different peak separations of different spectra.
The observed asymmetric evolution of the two components disfavors a simple AGN-driven outflow and a rotating disk-like NLR scenario.
Of the models considered, only a more complex non-axisymmetric rotating NLR provides a self-consistent explanation for the observed aperture-dependent variations. In this scenario, the blueshifted emitting region is more compact and remains largely enclosed by the smaller aperture, resulting in an increased line width with little change in flux. In contrast, the redshifted emitting region is more spatially extended and is therefore more strongly affected by the reduced aperture, leading to pronounced decreases in both flux and line width.
Our results illustrate the diagnostic potential of combining spectroscopic observations from different surveys with varying apertures to investigate the physical origin of DPNELs, while highlighting the challenges of distinguishing between different physical scenarios.
}

\maketitle

\section{Introduction} 
Active galactic nuclei (AGNs) commonly exhibit strong narrow emission lines originating from the narrow-line region (NLR), which extends from hundreds of parsecs to kiloparsecs \citep{Be02,Li13,He14,Hu22,Le25}.
A significant fraction of AGNs show double-peaked narrow emission lines (DPNELs), particularly in the \oiii$\lambda\lambda4959,5007$ doublet, indicating complex gas kinematics within the NLR \citep{Ge12,Ki20,Se21,Zh25,Zhang25}.
In addition to double-peaked profiles, some AGNs exhibit single narrow emission lines that are offset from the systemic velocity of their host galaxies \citep{Ba08,Co13,Zh23}. Such velocity offsets provide another observable signature of disturbed or noncircular gas motions in the NLR.
In the dual scenario, the offset can arise when only one core hosts an AGN and produces detectable narrow emission lines and the other does not show emission-line activity, resulting in a single velocity-offset component \citep{Co15}.
In the outflow scenario, the velocity offset can arise from NLR gas moving along the line of sight; emission from the receding side of the outflow is obscured, while the approaching side is more readily observed \citep{Zh26a}. These single-peaked velocity offsets and DPNELs may therefore represent different manifestations of disturbed NLR kinematics, with the latter providing additional kinematic information through the presence of two resolved velocity components.

For DPNELs, two major classes of models have been proposed to explain the origin of double-peaked narrow emission profiles. One scenario attributes the double peaks to the orbital motion of two distinct AGNs within a merging galaxy, each hosting its own NLR \citep{Zh04,Co15,Zh24}.
\cite{Ch25} interpreted the DPNELs shifted in the same direction using a gravitational acceleration model within the framework of a kiloparsec-scale AGN–galaxy pair, where an AGN and a non-emission-line galaxy orbit each other.
The other scenario explains the DPNELs as resulting from NLR kinematics associated with a single AGN, such as AGN-driven outflows \citep{Liu10,Co18,Ru19} or a rotating disk-like NLR \citep{Xu09,Sm12}. Disentangling these possibilities is crucial for understanding AGN fueling, feedback, and coevolution with their host galaxies, yet remains challenging because different models can often produce similar line profiles.

A key limitation of previous studies is that they relied largely on single-aperture spectra, which inherently average over the spatially extended NLR. Consequently, follow-up observations are essential to verify whether DPNELs are genuinely associated with dual AGN systems. Previous attempts to differentiate the dual AGN interpretation from alternative kinematic origins have primarily relied on high-resolution imaging and spatially resolved spectroscopic observations. In many cases, these two approaches have been used together to provide a more complete characterization of galaxies exhibiting DPNELs \citep{Liu10b,Sh11,Fu12,Mc15}.

Nevertheless, \citet{Vi15} demonstrated that even the combination of deep multiband imaging and long-slit spectroscopy does not always provide an unambiguous confirmation of dual AGNs in systems with double-peaked \oiii. In addition, although these observational techniques are effective for detailed investigations of individual candidates, they are observationally expensive and therefore inefficient for constructing large samples of dual AGN candidates selected from DPNEL galaxies. Such large statistical samples are, however, crucial for understanding the physical mechanisms responsible for the origin of DPNELs.

A more accessible but under-explored approach is to examine how the double-peaked line profile changes as a function of the aperture size used to extract the spectrum. 
Since the NLR is spatially extended and often exhibits significant kinematic stratification, and its characteristic dynamical timescales are typically much longer than decades \citep{Pe13}, the observed line profiles can depend strongly on the aperture size and spatial sampling. Different spectroscopic surveys employ substantially different aperture sizes, which correspond to different physical scales and can therefore probe different regions of the NLR.
Aperture-dependent spectral variability may therefore provide a powerful diagnostic of the physical origin of DPNELs. For example, dual AGNs, outflows, and rotating disks are expected to exhibit different responses to changes in the spatial sampling of the two components, particularly in terms of line width, flux, and peak separation.

We investigate the aperture dependence of DPNELs by comparing spectra obtained from different spectroscopic surveys with substantially different aperture sizes. 
Large spectroscopic surveys often provide multiple observations of the same target obtained with instruments employing different aperture diameters, enabling a direct investigation of aperture effects on the observed emission-line profiles.
Specifically, the aperture diameters are $1.5^{\prime \prime}$ for the Dark Energy Spectroscopic Instrument (DESI; \citealt{De19}), $3^{\prime \prime}$ for the Sloan Digital Sky Survey (SDSS; \citealt{Ah20}), and $3.3^{\prime \prime}$ for the Large Sky Area Multi-object Fiber Spectroscopic Telescope (LAMOST; \citealt{Cu12,Zh12}).
By analyzing the aperture-dependent variability of the blueshifted and redshifted components, we aim to distinguish between the dual AGNs, outflows, and rotating disk-like NLR models, and to determine which provides the most plausible explanation for the observed DPNELs.
Our approach provides a new, cost-effective diagnostic that leverages existing large-scale spectroscopic surveys and offers clues to the physical origin of DPNELs.

This paper is organized as follows. Section 2 presents the main hypotheses regarding the effect of aperture size. Section 3 describes the corresponding fitting results of the spectra. Section 4 discusses possible physical models. Section 5 provides a summary and our conclusions.
We have adopted the cosmological parameters $H_{0}=70 {\rm km/s/Mpc}$, $\Omega_{\Lambda}=0.7,$ and $\Omega_{\rm m}=0.3$.
\section{Main hypotheses}
Because different apertures probe different spatial extents of the NLR, aperture effects can influence the observed profiles of DPNELs. By examining variations in the line profiles obtained with different apertures, it is possible to constrain the physical origin of these features. At present, three main scenarios have been proposed to explain DPNELs: dual AGN systems, AGN-driven outflows, and rotating disk-like NLRs.

In the dual AGN scenario, the two NLRs are spatially distinct rather than forming a single continuous structure. As a result, changing the observational aperture effectively alters the relative sampling weights of these two independent emission regions.
Within the framework of the dual AGN scenario, it is essential to distinguish between two geometric configurations: one where the two nuclei are located in the same plane perpendicular to the line of sight, and another where they are aligned along the line of sight.

In the case where the two AGNs are spatially separated on the plane of the sky, changing the aperture size effectively modifies the relative sampling of the two NLRs.
Because the aperture is generally centered on the nucleus of the primary galaxy, a reduced aperture that still encompasses part of the NLR of the companion galaxy, so that DPNELs remain detectable, would also cover most of the NLR of the primary galaxy.
Spatially resolved spectroscopic studies have demonstrated that the kinematics of the NLR vary systematically with radius. In particular, the ionized gas closer to the nucleus typically exhibits higher velocity dispersion, giving rise to broader emission lines, whereas gas at larger distances produces narrower lines \citep{Da05,Ru05,De22}. For the primary component, a larger aperture integrates emission from both inner high-dispersion gas and outer low-dispersion gas, leading to a narrower line profile due to dilution by the outer regions.
Although a smaller aperture still encompasses most of the NLR, it excludes a larger fraction of the outer, lower-dispersion gas than of the inner, higher-dispersion gas. Consequently, the relative contribution of the inner region to the integrated emission increases, potentially broadening the observed line profile.
Meanwhile, since most of the NLR associated with the primary galaxy remains within the aperture, the emission-line flux from this region is expected to change only modestly.
For the secondary component, only the outer regions of the NLR that lie closer to the primary galaxy are more likely to be captured by a smaller aperture. As a result, the observed emission lines may appear narrower and exhibit lower line flux.
Since the observed peak separation traces the projected orbital motion along the line of sight, the peak separation between the two peaks remains unaffected by the change in aperture size.
If the space separation between the two nuclei is sufficiently large, a smaller aperture may fail to encompass the NLR of the companion galaxy altogether, such that only single-peaked narrow emission lines originating from the NLR of the primary galaxy are observed. Furthermore, because the smaller aperture preferentially samples the inner regions of the NLR, the resulting single-peaked emission lines may exhibit a larger line width and a lower line flux.

When the two AGNs are aligned along the line of sight, their projected separation on the plane of the sky is negligible. As a result, varying the aperture size does not significantly alter the relative sampling of the two NLRs.
As the aperture size decreases, the line widths of the two components tend to increase, whereas their fluxes are expected to decrease.
The peak separation more directly corresponds to their intrinsic relative motion along the direction of observer. Consequently, the peak separation between the two peaks remains nearly unchanged.

If the DPNELs originate from AGN-driven outflows or rotating disk-like NLRs, then changing the observational aperture effectively corresponds to re-weighting emission from different regions within a single, continuous kinematic structure. In the outflow scenario, this reflects variations in the relative contribution of gas at different radii within a biconical flow, while in the rotating disk scenario it corresponds to integrating a symmetric velocity field over different spatial extents. In both cases, because the emission arises from a coherent and continuous structure rather than from spatially distinct regions, the resulting spectral variations are expected to be largely symmetric and smooth.

In the outflow framework, the blueshifted component arises from the approaching side of the outflow (moving toward the observer), while the redshifted component originates from the receding side (moving away from the observer). Because the outflow constitutes a coherent and continuous kinematic structure, the two components trace the same physical system and differ only in their line-of-sight velocities. As a result, their line widths are expected to exhibit similar trends, such that both components either broaden or narrow in response to changes in aperture.
Because the velocity dispersion of the gas is larger at smaller distances from the nucleus, when a smaller aperture is used, it predominantly collects light from the inner, higher-dispersion regions, leading to broader observed narrow emission line profiles.
A reduction in aperture size decreases the spatial region of the NLR that is sampled, leading to a simultaneous decrease in the flux of both components.
Spatially resolved observations have shown that the outflow velocity in AGN NLRs can exhibit a variety of radial behaviors, including continuous acceleration, continuous deceleration, or acceleration followed by deceleration at larger radii. 
Spatially resolved observations indicate that the outflow velocity rises with radius up to a certain distance, after which it decelerates toward the systemic velocity \citep{Cr00,Da06,Fi13}. Consequently, reducing the aperture size can lead to different observational outcomes depending on the underlying velocity field. If the outflow accelerates with radius, a smaller aperture preferentially sampling the inner NLR will yield a smaller peak separation. Conversely, if the outflow decelerates outward, the smaller aperture will result in a larger peak separation. If the velocity field is approximately flat over the sampled region, the peak separation should remain unchanged.

In the rotating disk-like NLR scenario, the blueshifted component arises from the approaching side of the rotating disk, while the redshifted component originates from the receding side. 
A smaller aperture preferentially samples the central regions, where the velocity gradient is larger, leading to slightly broader line widths. In contrast, a larger aperture includes emission from a wider range of radii, mixing regions with different velocities and effectively diluting the observed line widths. Consequently, the blueshifted and redshifted components are expected to respond in a similar manner to changes in aperture, with their line widths decreasing in larger aperture.
In terms of flux, reducing the aperture size decreases the total emitting area being sampled. Since the two components originate from the same structure, their fluxes are expected to decrease simultaneously, while their flux ratio remains approximately constant.
The spatially resolved \oiii~kinematics presented by \citet[see Figs. 5 and 6]{Re25} showed that, in NLRs dominated by rotational motion, the velocity offset generally increases with projected distance from the center. Under such a velocity field, a smaller aperture preferentially samples the inner regions of the NLR, where the rotational velocities are lower than those at larger radii. Consequently, the peak separation measured from a smaller-aperture spectrum may be smaller than that derived from a larger-aperture spectrum.

The aperture-dependent behaviors expected under different physical scenarios are summarized in Table~\ref{tab2}. As discussed above, the predicted responses of the line widths, line fluxes, and the peak separations to aperture changes differ among the dual AGN, AGN-driven outflow, and rotating disk-like NLR models when the aperture size changes. These distinct predictions provide a useful framework for evaluating the physical origin of the DPNELs.

\section{Data and methods}
\begin{figure*}
        \centering\includegraphics[width=5.8cm,height=3.6cm]{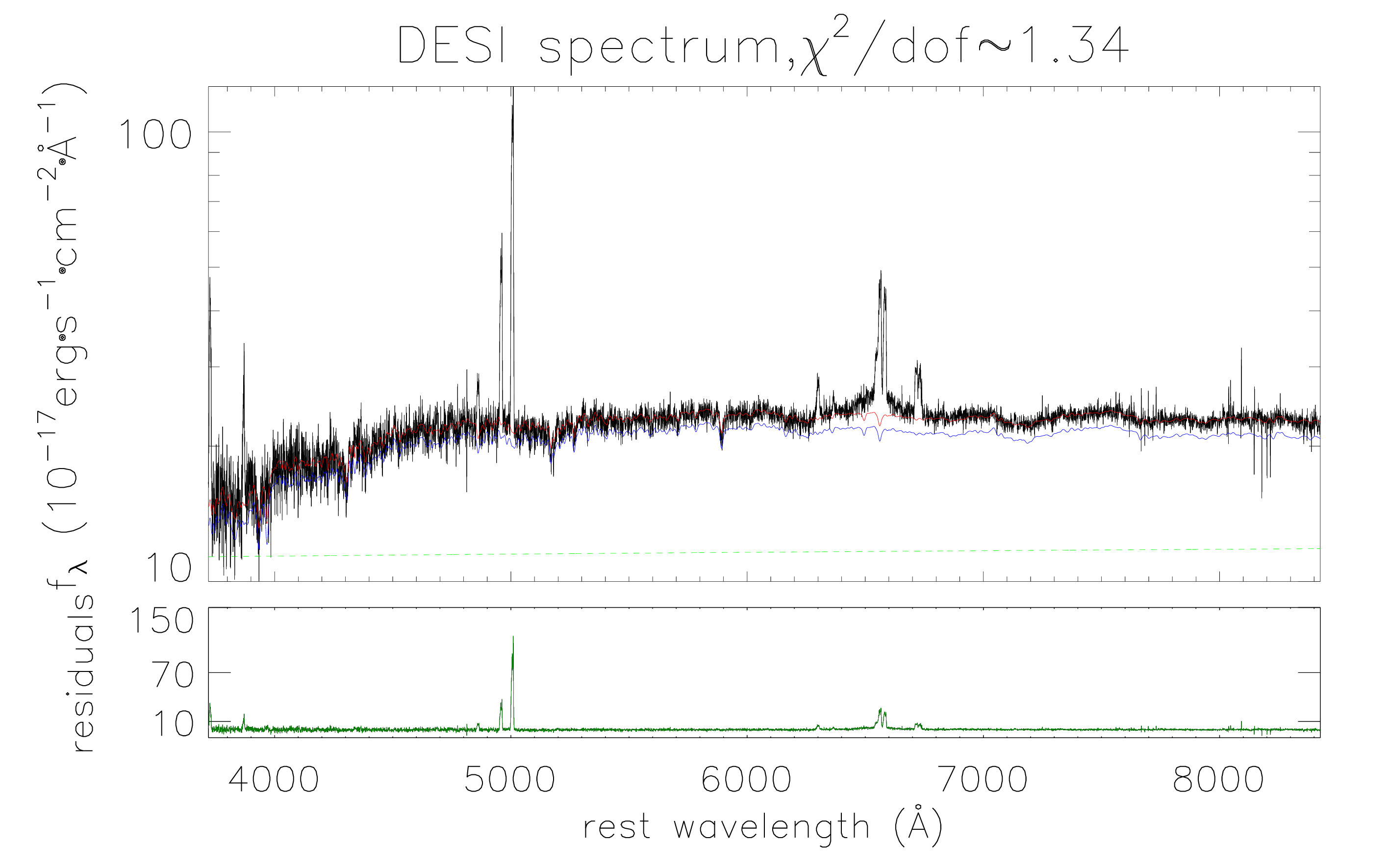}
        \centering\includegraphics[width=5.8cm,height=3.6cm]{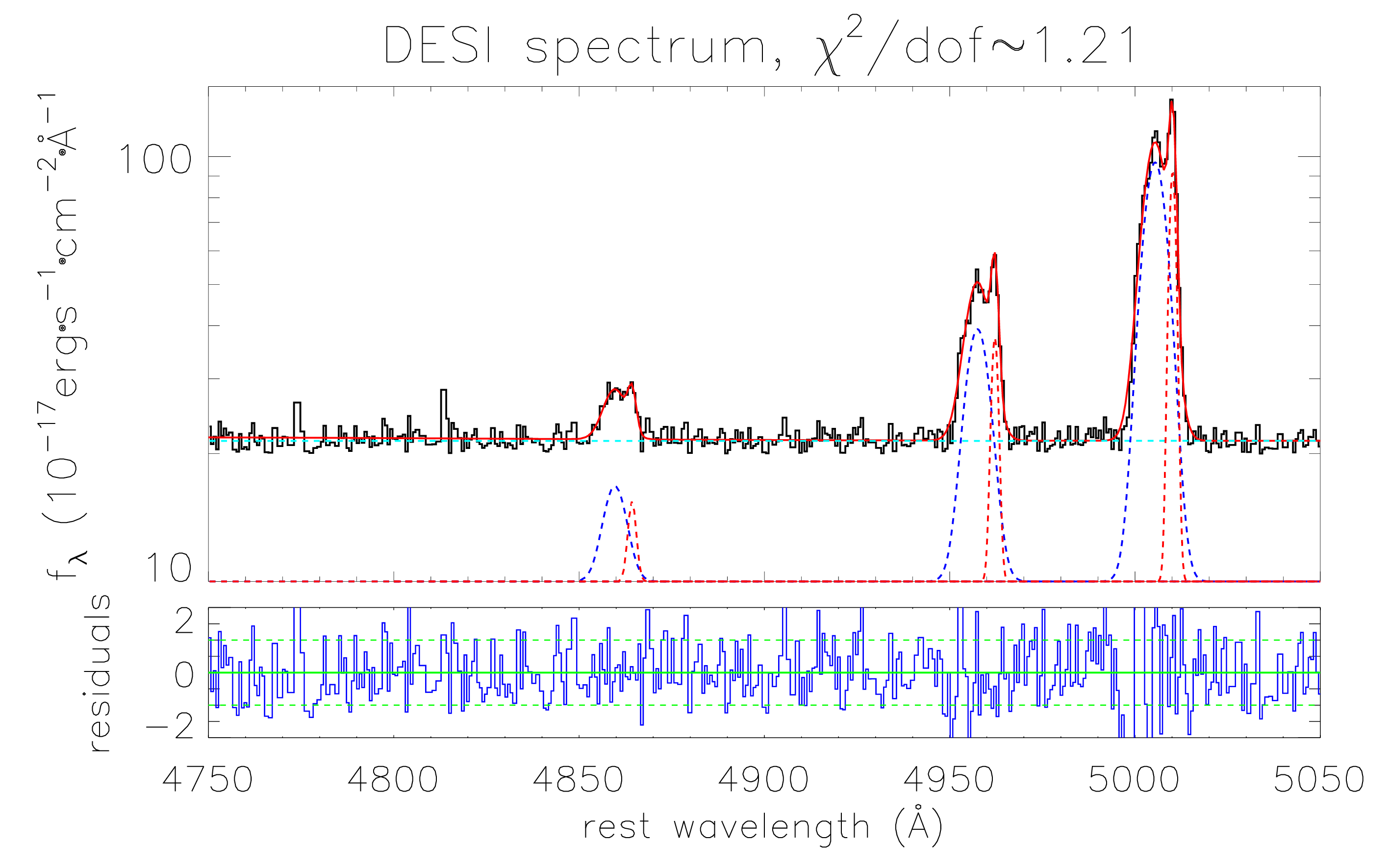}     
        \centering\includegraphics[width=5.8cm,height=3.6cm]{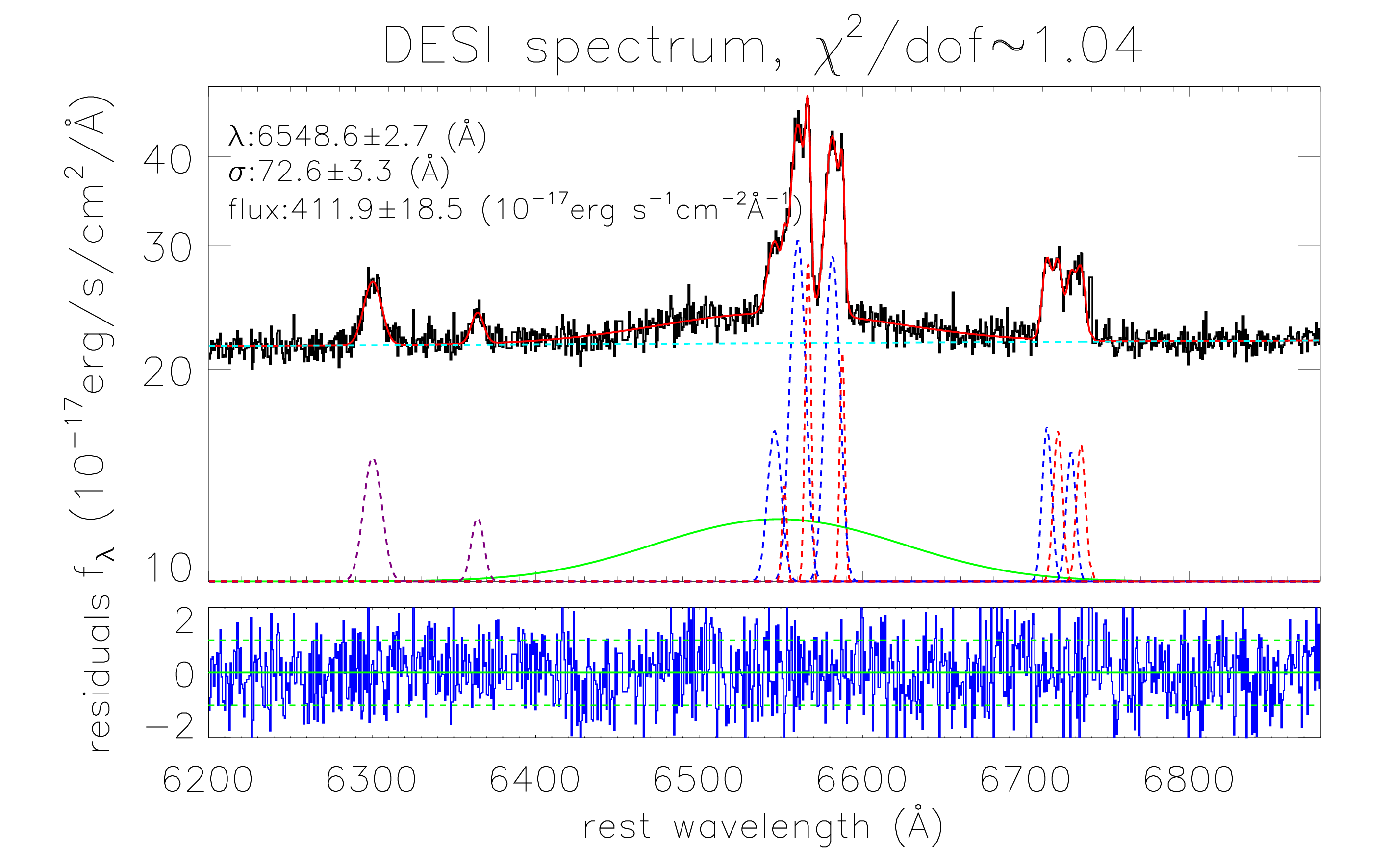}       
        \centering\includegraphics[width=5.8cm,height=3.6cm]{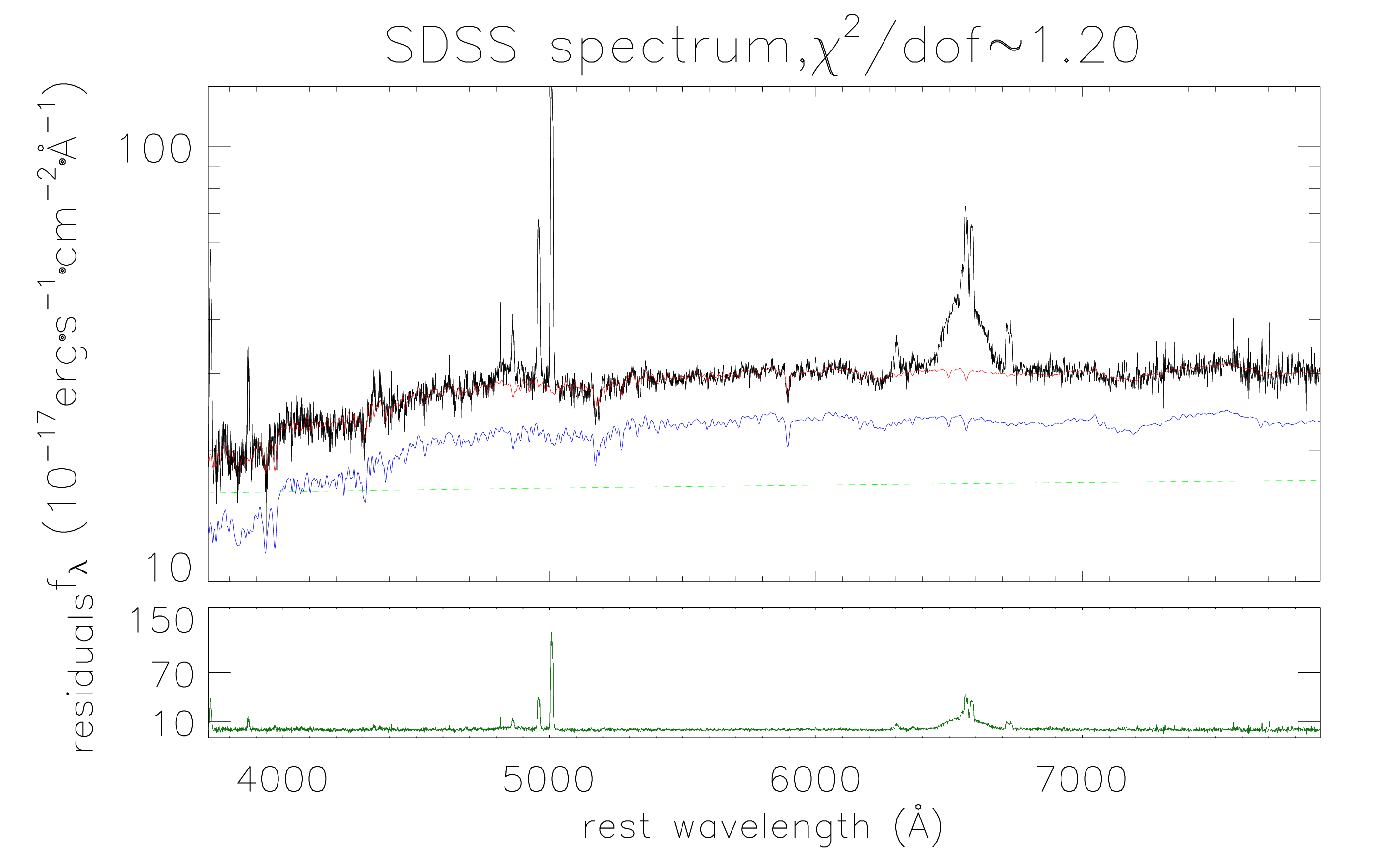}     
        \centering\includegraphics[width=5.8cm,height=3.6cm]{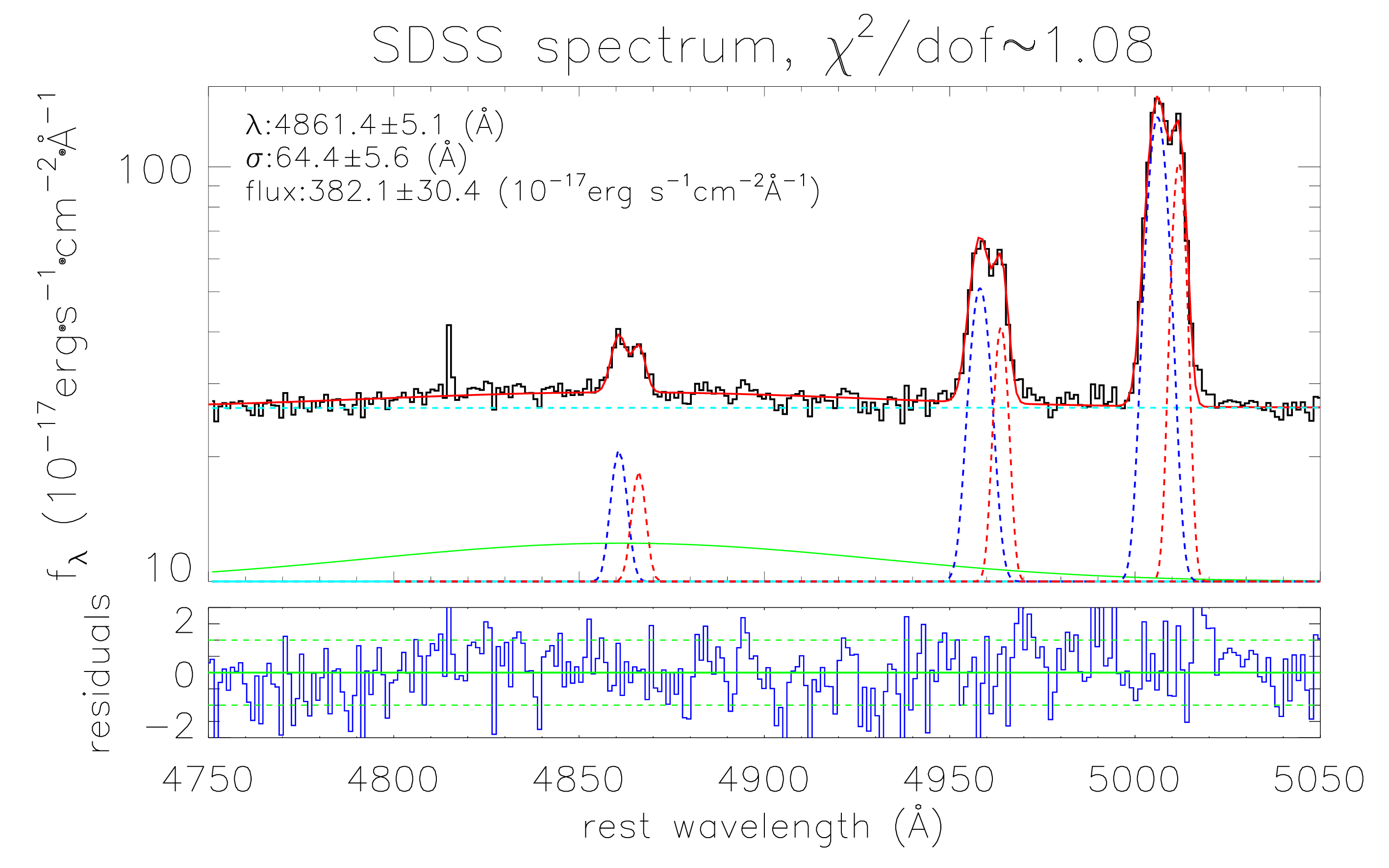}     
        \centering\includegraphics[width=5.8cm,height=3.6cm]{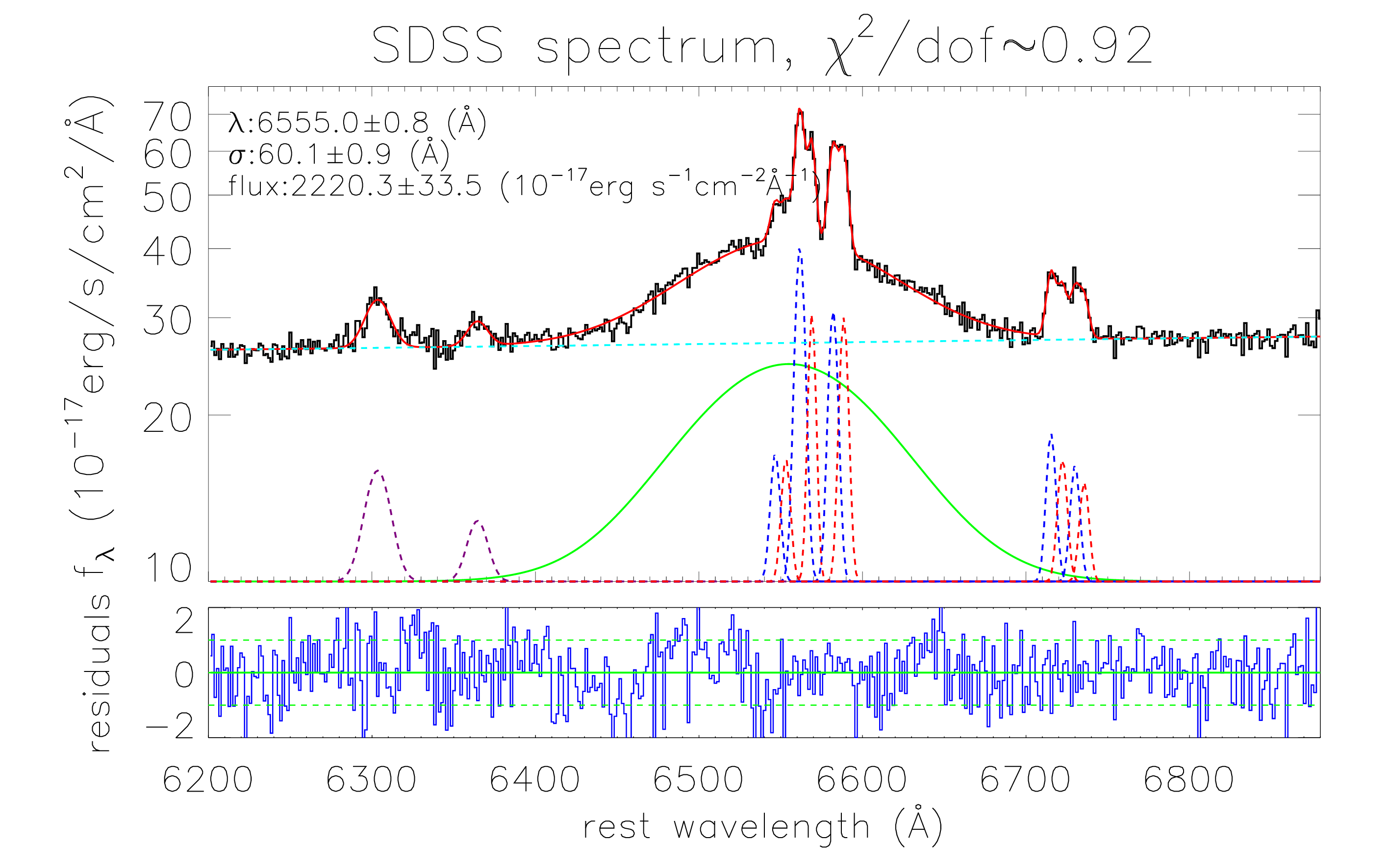}
        \centering\includegraphics[width=5.8cm,height=3.6cm]{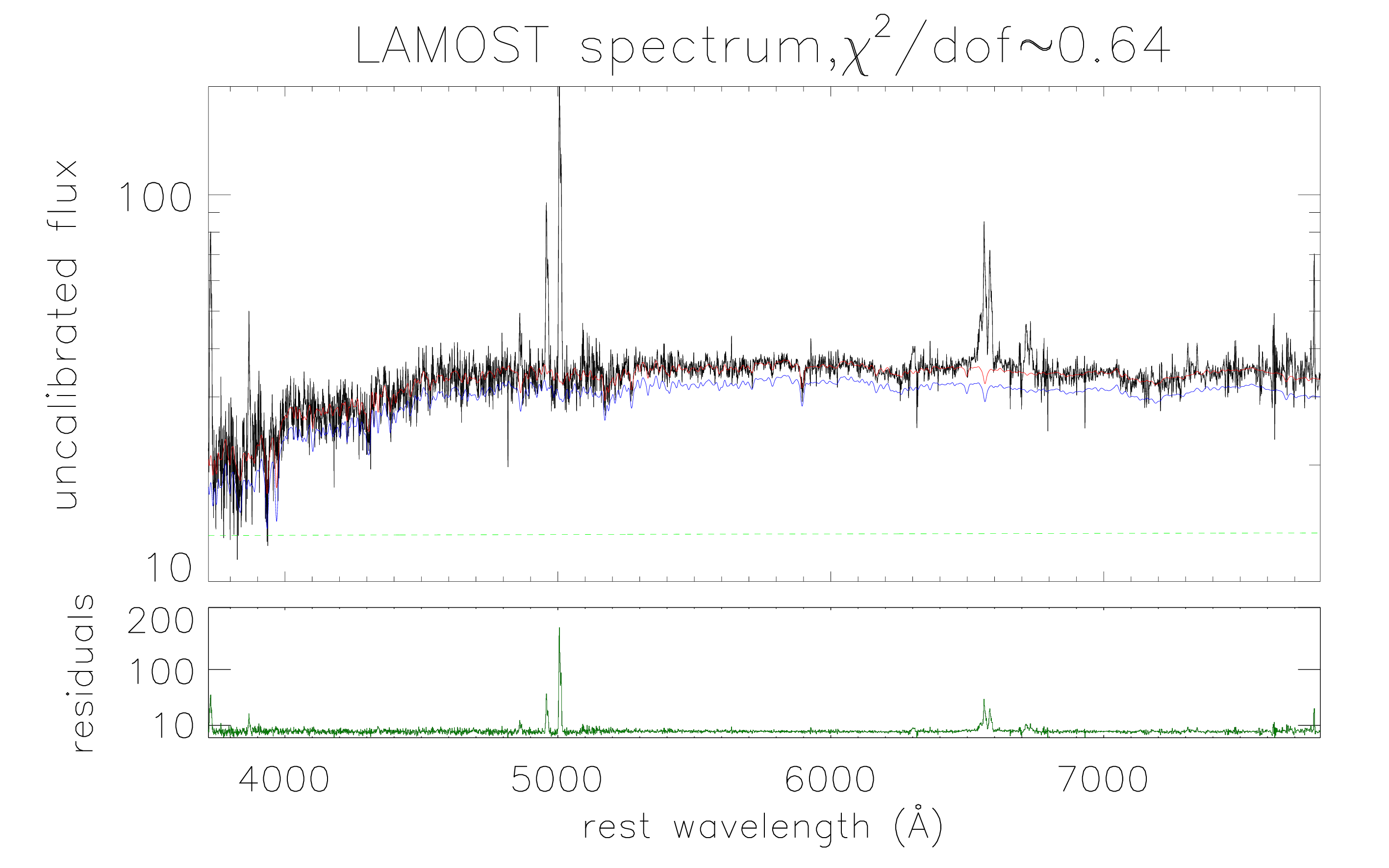}
        \centering\includegraphics[width=5.8cm,height=3.6cm]{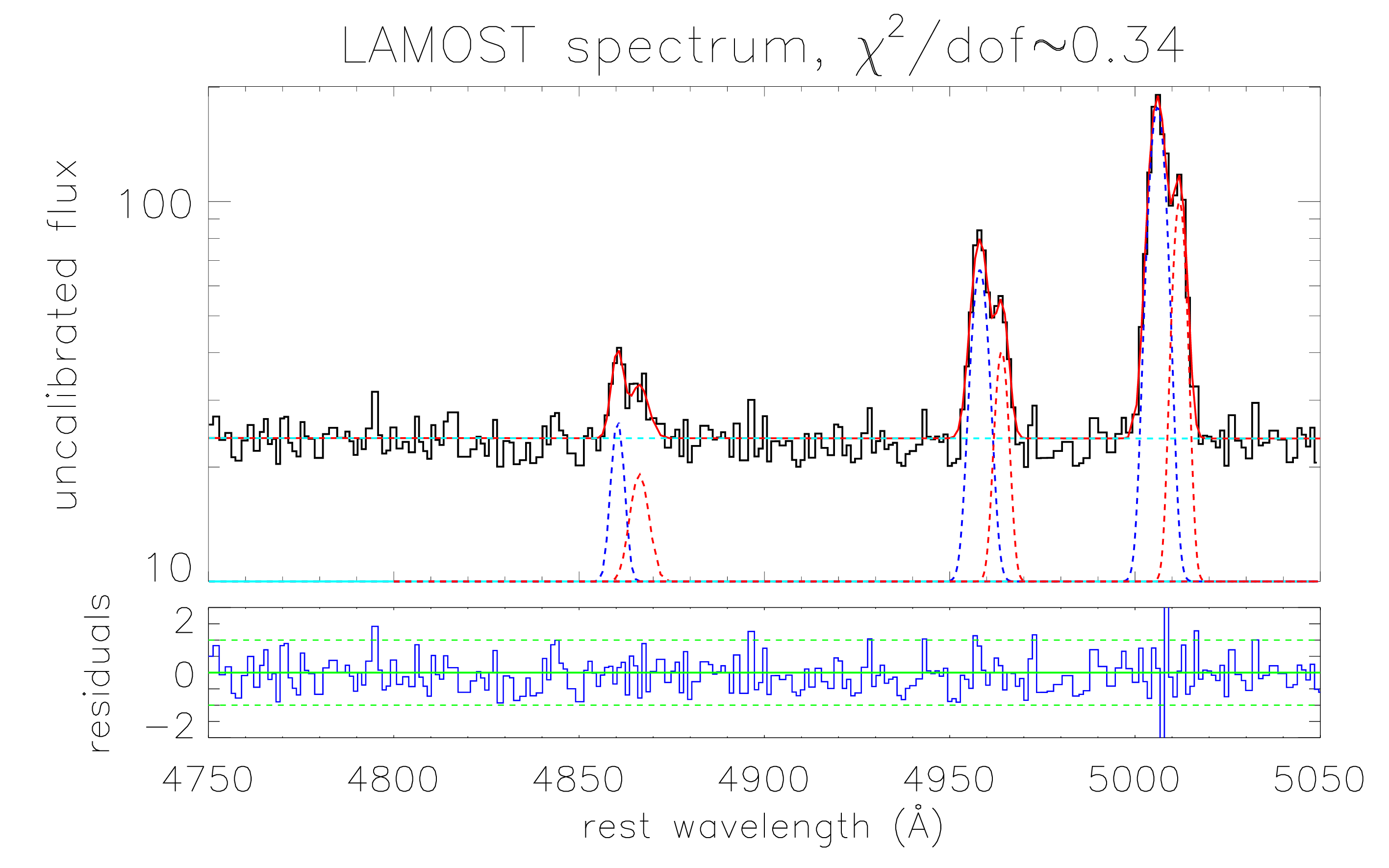}   
        \centering\includegraphics[width=5.8cm,height=3.6cm]{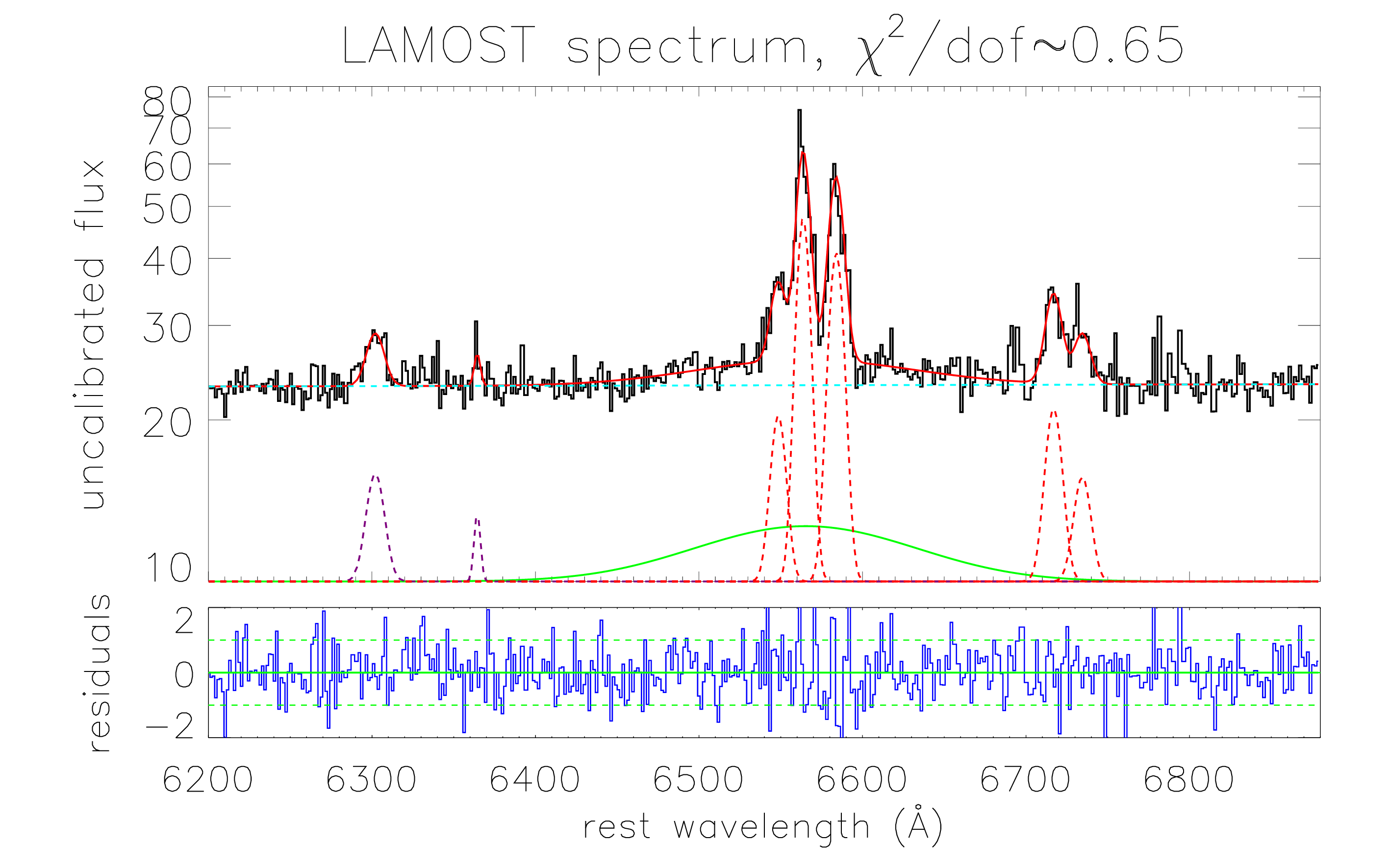}
\caption{Spectra and corresponding fitting results of SDSS J0941 from DESI (top row), SDSS  (middle row), and LAMOST (bottom row).
In each upper panel, the solid red line shows the best fitting results, with $\chi^2/\rm dof$ shown in the title.
\textit{Left column}: In the upper panels, all lines have been shifted upward by 10 units for clarity. The solid black lines represent observed spectra, the solid blue lines show starlight, and the dashed green lines show the AGN continuum emission. In the lower panels, the solid green lines show the line spectra calculated by the observed spectra minus the best fitting results. 
\textit{Middle and right columns}: In the upper panels, the line spectra, the best fitting results, and continuum emission have been shifted upward by 20 units, and the Gaussian functions have been shifted upward by 10 units. The solid black lines show the line spectra after subtracting starlight.\ The solid green lines show broad H$\beta$ and H$\alpha$ emission lines, with their corresponding central wavelengths $\lambda$(\AA), line widths $\sigma$(\AA), and fluxes ($10^{-17}{\rm erg/s/cm^{2}}$) listed in the panels. The dashed blue and red lines represent blueshifted and redshifted components of narrow H$\beta$ and H$\alpha$, \oiii, 
[N~{\sc ii}], and [S~{\sc ii}] doublets, the dashed purple lines show the [O~{\sc i}] doublet, and the dashed cyan lines show continuum emission.
In the lower panels, the solid blue lines represent the normalized residuals, and the solid and dashed green lines show residuals=0 and $\pm$1, respectively.           
}
\label{fig1}
\end{figure*}

To evaluate the applicability of the proposed method, we selected an object exhibiting relatively prominent variations in its observed DPNEL profiles across spectra obtained by different surveys. 
To identify such an object, we cross-matched the sample of 62 type 1 AGNs with DPNELs identified from SDSS DR16 by \citet{Zh25}, against the DESI spectroscopic data. Among the matched objects, SDSS J0941 was the only object for which we identified relatively prominent variations in the DPNEL profiles. We therefore selected this object as a test case to investigate how the proposed diagnostic method performs when applied to multi-epoch spectra with different observational characteristics. In addition to the SDSS spectrum obtained on MJD = 51911 and the DESI spectrum obtained on MJD = 59628, SDSS J0941 was observed by LAMOST on MJD = 57039. 
The spectral resolution $R$ differs among the three surveys and varies with wavelength. At the observed wavelength of the \oiii~line in SDSS J0941, the resolving powers of the SDSS, LAMOST, and DESI spectra are approximately $R_{\rm SDSS}$=1844, $R_{\rm LAMOST}$=1604, and $R_{\rm DESI}$=3080, respectively. These correspond to instrumental velocity dispersions of approximately $\sigma_{\rm inst}$=69 km/s, 79 km/s, and 41 km/s, respectively.
The spectra from all three surveys are presented in the left panels of Fig. \ref{fig1}.

The NLR is an extended emission region with a typical size of hundreds of parsecs to several kiloparsecs. Owing to its large size, significant variations in the overall NLR luminosity are generally not expected on timescales of only a few decades \citep{Pe13}. Consequently, the $\sim$20-year interval between the LAMOST, SDSS, and DESI observations is unlikely to have produced substantial intrinsic changes in the global structure or kinematic properties of the NLR.
Therefore, the differences observed in the narrow emission-line profiles are more plausibly attributed to aperture effects arising from the different aperture sizes of the three spectroscopic surveys.

\setlength{\tabcolsep}{4pt}
\begin{table*}
        \caption{Features of the main narrow emission lines of SDSS J0941.}
        \begin{center}
                \begin{tabular}{ccccc|cccc}     
                        \hline\hline
                        Line& $\lambda_{\rm SDSS}$ & $\sigma_{\rm SDSS}$  &  $\rm flux_{\rm SDSS}$  & $\Delta\upsilon_{\rm SDSS}$& $\lambda_{\rm DESI}$ & $\sigma_{\rm DESI}$  &  $\rm flux_{\rm DESI}$  & $\Delta\upsilon_{\rm DESI}$ \\ \hline
                        \multirow{2}{*}{H$\beta$}       & 4860.7$\pm$0.4   & 1.9$\pm$0.3 & 50.0$\pm$8.7 & \multirow{2}{*}{327.0$\pm$49.4}        & 4859.7$\pm$0.6 & 3.1$\pm$0.6 & 53.2$\pm$9.0 & \multirow{2}{*}{283.8$\pm$49.4} \\ 
                        { }     & 4866.0$\pm$0.4   & 1.7$\pm$0.4 & 35.6$\pm$8.0 & { } & 4864.3$\pm$0.2  & 1.1$\pm$0.3 & 14.1$\pm$7.0 & { } \\ 
                        \hline
                        \multirow{2}{*}{[O \sc{iii}]}   &5006.1$\pm$0.1  & 2.6$\pm$0.1 & 776.8$\pm$20.3  & \multirow{2}{*}{341.4$\pm$12.0}       &5005.4$\pm$0.1  & 3.3$\pm$0.1 & 727.1$\pm$10.0  & \multirow{2}{*}{281.5$\pm$12.0}\\ 
                        { }     & 5011.8$\pm$0.1  & 1.7$\pm$0.1 & 403.9$\pm$18.7  & {  } &5010.1$\pm$0.1  & 1.0$\pm$0.1 & 195.5$\pm$7.4& {  } \\ 
                        \hline  
                        \multirow{2}{*}{H$\alpha$}      & 6561.7$\pm$0.3 & 2.9$\pm$0.2 & 223.9$\pm$18.5 & \multirow{2}{*}{338.2$\pm$32.0} & 6560.5$\pm$0.3 & 4.0$\pm$0.2 & 204.3$\pm$11.5 & \multirow{2}{*}{283.3$\pm$18.3}\\ 
                        { }     & 6569.1$\pm$0.4  & 2.3$\pm$0.3 & 116.6$\pm$17.6 & { } & 6566.7$\pm$0.1  & 1.5$\pm$0.1 & 70.5$\pm$9.7 & { } \\ 
                        \hline  
                        \multirow{2}{*}{[N \sc{ii}]}    & 6582.0$\pm$0.6 & 2.8$\pm$0.4 & 147.8$\pm$29.0 & \multirow{2}{*}{296.1$\pm$36.4} & 6581.5$\pm$0.2 & 3.9$\pm$0.2 & 186.5$\pm$7.8 & \multirow{2}{*}{277.9$\pm$13.7}\\ { } 
                        & 6588.5$\pm$0.6 & 2.7$\pm$0.4 & 133.2$\pm$28.9 & { } & 6587.6$\pm$0.1 & 1.4$\pm$0.1 & 38.7$\pm$6.2 & { }\\ 
                        \hline
                        \multirow{2}{*}{[S \sc{ii}]}    & 6715.3$\pm$0.8 & 2.7 & 58.9$\pm$16.6& \multirow{2}{*}{312.6$\pm$75.9} & 6712.8$\pm$0.4 & 2.7 & 43.9$\pm$5.9& \multirow{2}{*}{299.2$\pm$31.3}\\
                        { }     & 6722.3$\pm$0.9  & 2.8 & 46.2$\pm$19.7 & { } & 6719.5$\pm$0.3  & 2.5 & 40.5$\pm$6.4 & { }\\
                        \hline
                        \multirow{2}{*}{[S \sc{ii}]}    & 6730.0$\pm$1.5  & 2.8$\pm$0.6 & 43.2$\pm$24.0 & \multirow{2}{*}{249.5$\pm$133.7} & 6727.5$\pm$0.4 & 2.7$\pm$0.3 & 35.4$\pm$6.8 & \multirow{2}{*}{271.8$\pm$26.7}\\
                        { }     & 6735.6$\pm$1.5  & 2.8$\pm$0.9 & 35.5$\pm$ 20.3 & { } & 6733.6$\pm$0.4  & 2.6$\pm$0.3 & 36.2$\pm$ 5.9 & { }\\ \hline
\end{tabular}
\tablefoot{
For each emission line, the table lists the central wavelength $\lambda$ (\AA), line width $\sigma$ (\AA), line flux ($10^{-17}{\rm erg/s/cm^{2}}$), and the peak separation $\Delta\upsilon$ (km/s) between the blueshifted and redshifted components. The subscripts ``SDSS'' and ``DESI'' denote the corresponding measurements obtained from the SDSS and DESI spectra, respectively.                     
}\label{tab1}
\end{center}    
\end{table*}

\begin{figure}
\centering\includegraphics[width=8cm,height=7cm]{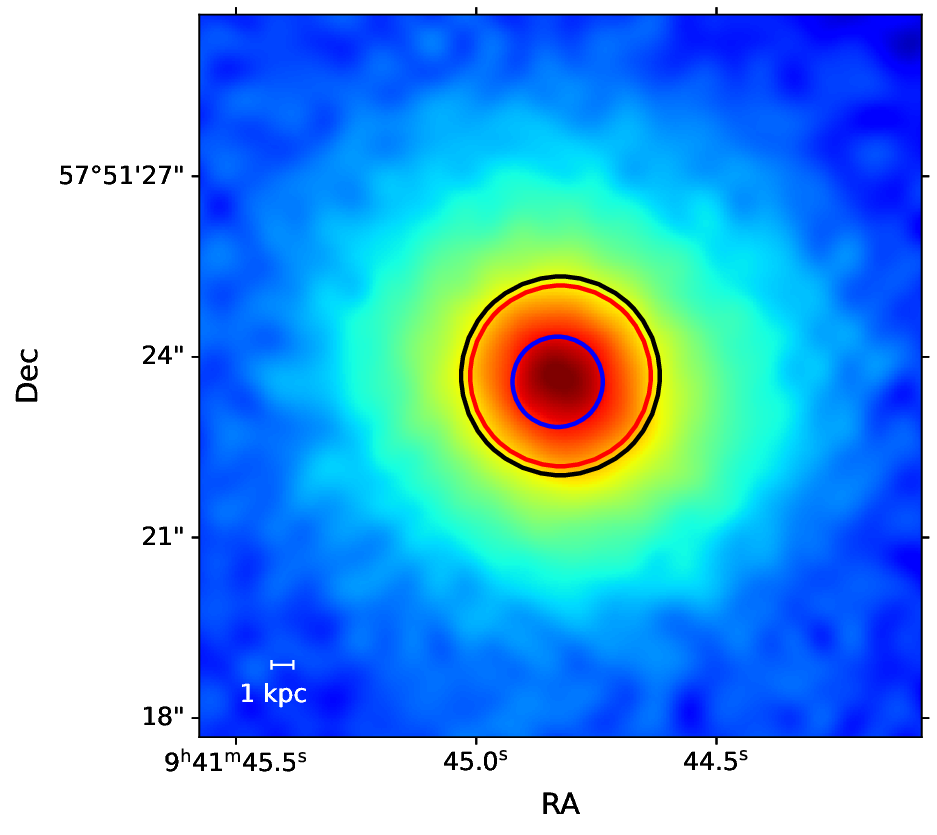}        
\caption{Image of SDSS J0941 in the g band from DESI. The plotting scale of 1 kpc is shown in white. The blue circle represents the aperture scope of the DESI spectrum, the red circle represents the aperture scope of the SDSS spectrum, and the black circle represents the aperture scope of the LAMOST spectrum.
}
\label{fig2}
\end{figure}

The center of the SDSS aperture is located at \texttt{plug\_ra = 145.43677} and \texttt{plug\_dec = 57.856579}, while the center of the LAMOST aperture is at \texttt{RA\_OBS = 145.436770} and \texttt{DEC\_OBS = 57.856579}, indicating that the LAMOST aperture position is fully consistent with that of the SDSS observation. The DESI aperture is centered at \texttt{MEAN\_FIBER\_RA = 145.4367943889} and \texttt{MEAN\_FIBER\_DEC = 57.8565516247}. These positions are also highly consistent, with a positional offset of only 0.036 arcsec. The corresponding aperture scope are illustrated in Fig.~\ref{fig2}.

Following the method of \citet{Zh25}, we modeled the host-galaxy starlight using the simple stellar population (SSP) method \citep{Br93,Ci05,ca17}. The stellar component was represented by 39 SSP templates \citep{Ka03,Br03} and the AGN continuum was described by a power-law component. The observed spectra were fitted with the combined SSP and power-law model using the Levenberg-Marquardt least-squares minimization technique \citep{Ma09}.
The fitting results for observed spectra of SDSS J0941 are shown in the left panels of Fig.~\ref{fig1}.

After subtracting the host galaxy contribution, the emission lines around H$\beta$ (rest-frame 4750–5050~\AA) and H$\alpha$ (6200–6880~\AA) were modeled. The underlying continuum was represented by a single power-law component. Each narrow emission line was fitted with two Gaussian components to account for the double-peaked profiles. The broad H$\beta$ and H$\alpha$ lines were each modeled with a single broad Gaussian function.
Parameter constraints were applied as follows. For the \oiii\ and [N~{\sc ii}] doublets, the central wavelengths and line widths of the blueshifted and redshifted components were tied to those of their respective companion lines within each doublet, with the flux ratios fixed at the theoretical value of 3:1. For the [S~{\sc ii}] doublet, the line widths of the blueshifted and redshifted components were required to be consistent with those of their corresponding companion lines in velocity space.

The main emission lines of SDSS J0941 were fitted using the Levenberg–Marquardt least-squares minimization method. The fitting results for these spectra are shown in the middle and right panels of Fig.~\ref{fig1}. 
The SDSS spectrum and the corresponding spectral fitting results of SDSS J0941 were previously presented in \citet{Zh25}. In this work, we adopted the same method to subtract the host-galaxy contribution. However, in \citet{Zh25}, only the \oiii~were modeled as DPNELs, whereas the other major narrow emission lines were fitted using single-Gaussian profiles. Here, we reanalyzed the SDSS spectrum and performed double-Gaussian fitting for all major narrow emission lines.
Similar as done in \citet{Zh25}, we initially included an additional wing component in the fitting of the \oiii~emission line. However, the fitted wing component in the SDSS spectrum has a flux lower than three times its uncertainty, suggesting that its detection is not statistically robust. Furthermore, no significant wing component was identified in both the DESI and LAMOST spectra.
Therefore, we did not consider the wing component to be reliable and excluded it from the final fitting model.

As shown in Fig. \ref{fig1}, SDSS J0941 can be classified as a changing-look AGN \citep{No18,Ri23,Pa24}, exhibiting a clear weakening of its broad Balmer emission lines. 
From 2001 to 2022, the spectrum shows a remarkable evolution: the broad H$\beta$ line gradually disappears, while the broad H$\alpha$ line significantly decreases in strength, marking a transition from a type 1 AGN (with prominent broad Balmer lines) to a type 1.9 AGN (where only a weak broad H$\alpha$ remains and broad H$\beta$ is no longer detectable).
However, the primary focus of this work is to investigate the spatial-scale dependence of the DPNELs. Therefore, we did not further discuss the physical origin of the changing-look phenomenon in this paper.
It should be noted that the relatively low spectral resolution and signal-to-noise ratio of the LAMOST spectrum make the double-peaked structures of the narrow H$\alpha$, [N~{\sc ii}], and [S~{\sc ii}] emission lines difficult to resolve clearly. Therefore, each of these emission lines was fitted with a single Gaussian component. 
Although instrumental resolution can contribute to the observed differences in line widths of double-peaked profiles, it is unlikely to fully account for the substantial variations observed among the different spectra.
Besides, since the LAMOST spectrum is not flux-calibrated, and its  aperture is comparable in size to that of the SDSS spectrum, we only performed a basic analysis of the LAMOST spectrum. Consequently, the emission-line parameters derived from the LAMOST spectrum are not included in Table~\ref{tab1}.
The derived parameters of the narrow emission lines from SDSS and DESI spectra are listed in Table~\ref{tab1}, where the peak separations and their uncertainties are calculated from the best-fit central wavelengths and their associated uncertainties.

\begin{figure}
\centering\includegraphics[width=7cm,height=7cm]{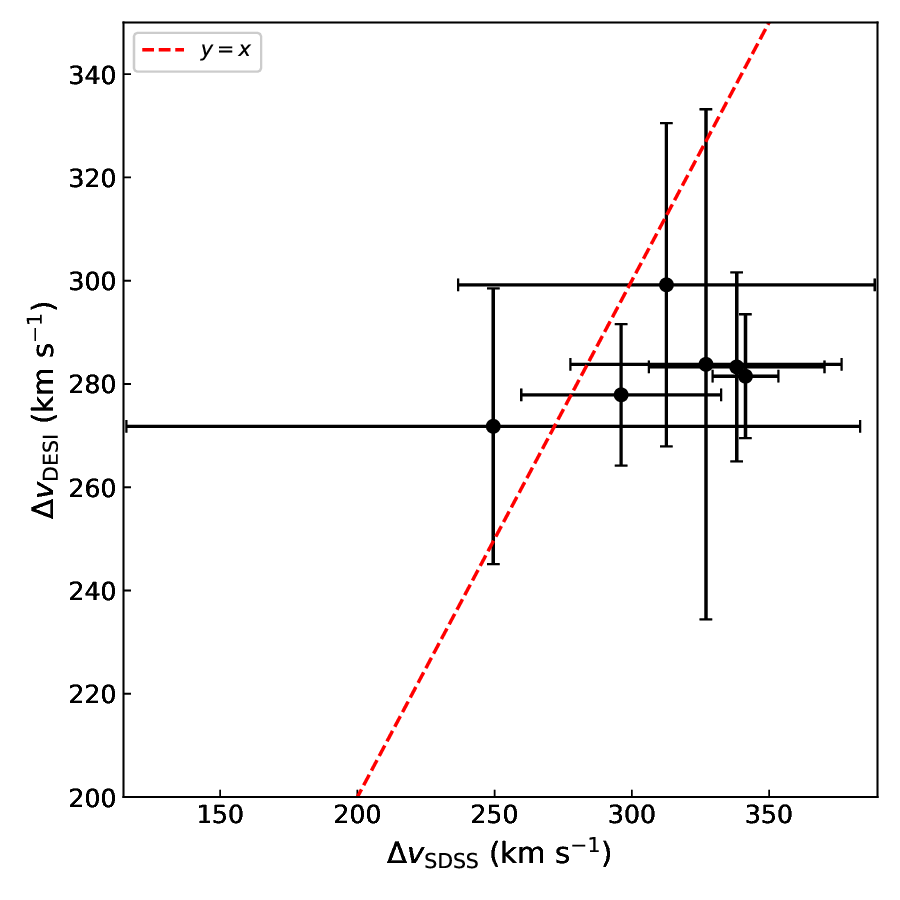}        
\caption{Comparison of the peak separations between the DPNEL components measured from the SDSS and DESI spectra. The dashed red line represents the one-to-one relation.
}
\label{fig3}
\end{figure}

Figure \ref{fig3} compares the peak separations of the DPNELs measured from the SDSS and DESI spectra. 
For five of the six emission lines considered, the peak separations derived from the DESI spectrum are smaller than those measured from the SDSS spectrum.
The mean absolute difference between the two measurements is approximately 28 km/s, corresponding to about 8\% of the typical peak separation ($\sim$ 340 km/s in SDSS spectrum).
The weighted mean difference is 56$\pm$14 km/s, corresponding to a significance of approximately $4\sigma$.
Given the measurement uncertainties, we therefore conclude that DESI measurements yield slightly lower peak separations compared to SDSS.
To minimize the possibility that the measured peak separation is affected by uncertainties in the spectral fitting, we also determined the peak positions directly from the host-subtracted spectra.
Specifically, after subtracting the stellar continuum model, we identified the two local maxima of the \oiii$\lambda5007$ emission-line profile and took their corresponding wavelengths as the peak positions. We then calculated the velocity separation between the two peaks from their wavelength difference. This direct measurement is independent of the multicomponent emission-line fitting and provides an independent consistency check.
The peak positions obtained in this way are consistent with those derived from the multicomponent spectral fitting.
This agreement demonstrates that the smaller velocity offset measured in the DESI spectrum relative to that in SDSS is a robust observational result.

\begin{table}
\centering
\caption{Comparison of the observed changes in the \oiii~line profiles induced by a decreasing aperture size with the predictions of different physical scenarios for the origin of the DPNELs.}
\label{tab2}
\begin{tabular}{llcccc}
                \hline
                Model &
                Quantity &
                Expected &
                Observed &
                Result \\
                \hline
                
                \multirow{6}{2cm}{Dual AGN 
                        (separated on the plane of the sky)}            
                & $\sigma_{b}$
                & $\uparrow$ ($\downarrow$ or 0)
                & $\uparrow$
                & $\checkmark$
                \\
                
                & $\sigma_{r}$
                & $\downarrow$ or 0 ($\uparrow$)
                & $\downarrow$
                & $\checkmark$
                \\
                
                & $f_{b}$
                & $\approx$ ($\downarrow$ or 0)
                & $\approx$
                & $\checkmark$
                \\
                
                & $f_{r}$
                & $\downarrow$ or 0 ($\approx$)
                & $\downarrow$
                & $\checkmark$
                \\
                
                & $\Delta\upsilon$
                & $\approx$ (0)
                & $\downarrow$
                & $\times$
                \\
                \hline
                
                \multirow{6}{2cm}{Dual AGN
                        (along the line of sight)}
                
                & $\sigma_{b}$
                & $\uparrow$
                & $\uparrow$
                & $\checkmark$
                \\
                
                & $\sigma_{r}$
                & $\uparrow$
                & $\downarrow$
                & $\times$
                \\
                
                & $f_{b}$
                & $\downarrow$
                & $\approx$
                & $\times$
                \\
                
                & $f_{r}$
                & $\downarrow$
                & $\downarrow$
                & $\checkmark$
                \\
                
                & $\Delta\upsilon$
                & $\approx$
                & $\downarrow$
                & $\times$
                \\
                \hline
                
                \multirow{6}{2cm}{AGN-driven outflow}
                
                & $\sigma_{b}$
                & $\uparrow$
                & $\uparrow$
                & $\checkmark$
                \\
                
                & $\sigma_{r}$
                & $\uparrow$
                & $\downarrow$
                & $\times$
                \\
                
                & $f_{b}$
                & $\downarrow$
                & $\approx$
                & $\times$
                \\
                
                & $f_{r}$
                & $\downarrow$
                & $\downarrow$
                & $\checkmark$
                \\
                
                & $\Delta\upsilon$
                & $\approx$($\downarrow$ or $\uparrow$)
                & $\downarrow$
                & $\checkmark$
                \\
                \hline
                
                \multirow{6}{2cm}{rotating disk-like NLR}
                
                & $\sigma_{b}$
                & $\uparrow$
                & $\uparrow$
                & $\checkmark$
                \\
                
                & $\sigma_{r}$
                & $\uparrow$
                & $\downarrow$
                & $\times$
                \\
                
                & $f_{b}$
                & $\downarrow$
                & $\approx$
                & $\times$
                \\
                
                & $f_{r}$
                & $\downarrow$
                & $\downarrow$
                & $\checkmark$
                \\
                
                & $\Delta\upsilon$
                & $\downarrow$
                & $\downarrow$
                & $\checkmark$
                \\
                \hline          
        \end{tabular}
\tablefoot{
The $b$ and $r$ components correspond to the blueshifted and redshifted narrow-line systems, respectively. $\sigma$ and $f$ denote the line width and integrated flux of each component, while $\Delta\upsilon$ refers to the peak separation between the two peaks. The symbols $\uparrow$, $\downarrow$, $\approx$, and 0 indicate an increase, a decrease, no significant change, and disappearance, respectively, as the aperture size decreases from the SDSS aperture to the DESI aperture. The Result column indicates whether the observed behavior is consistent ($\checkmark$) or inconsistent ($\times$) with the model predictions. 
}
\end{table}

\section{Discussions}\label{sec4}
In this section we investigate the origin of the DPNELs in SDSS J0941 by examining their spectral properties obtained with different apertures. Due to the lack of reliable flux calibration in the LAMOST spectrum, we limited its use to a qualitative discussion of the flux ratio between the two components of the double-peaked lines. The main analysis of aperture-dependent variations in the double-peaked narrow emission properties is therefore based on a comparison between the SDSS and DESI spectra.

The reduction in aperture size from the SDSS to the DESI observations leads to noticeable changes in the \oiii~line profiles. The blueshifted component becomes broader, whereas the redshifted component becomes narrower. Meanwhile, the flux of the blueshifted component is almost unchanged, while that of the redshifted component decreases significantly. In addition to the changes in the line profiles, the peak separation also decreases slightly.
Such distinct aperture-dependent behaviors are expected to differ among various physical scenarios proposed for the origin of DPNELs. Therefore, we compared the observed variations with the predictions of the dual AGN, outflow, and rotating-disk models. The results of this comparison, including the degree of agreement between the observations and each model prediction, are summarized in Table~\ref{tab2}.

In the dual AGN scenario, the two sets of narrow emission lines are interpreted as arising from the NLRs associated with two distinct active nuclei.  
As shown in Table 1, the emission-line ratios obtained from our spectral decomposition place both components in the AGN-dominated region of the Baldwin-Phillips-Terlevich \citep{Ke01,Ka03,Ch25} diagnostic diagrams.
Given the redshift of SDSS J0941 (z=0.159), the aperture diameters of SDSS and DESI correspond to physical scales of approximately 8.1 kpc and 4.0 kpc, respectively. 
When the two nuclei are located in the same plane perpendicular to the line of sight, given that the NLR typically extends over kiloparsec scales \citep{Po24}, the aperture of SDSS is sufficiently large to encompass the NLR of the primary galaxy as well as a substantial fraction of the NLR associated with the companion galaxy. In contrast, the smaller aperture of DESI samples a more limited spatial region, and therefore is likely to include a significantly smaller portion of NLR of the companion galaxy.
In this scenario, the blueshifted narrow emission-line component in SDSS J0941 is more likely associated with the primary galaxy, while the redshifted component originates from the companion galaxy.
A comparison between the spectra obtained by SDSS and DESI shows that the line width of the blueshifted component is narrower in the SDSS spectrum than in the DESI spectrum, whereas the redshifted component exhibits a broader line width in SDSS than in DESI. 
The flux of the blueshifted component, associated with the primary galaxy, changes only slightly between the SDSS and DESI spectra, while that of the redshifted component decreases noticeably in DESI spectra. 
This behavior can be naturally explained by aperture effects.

SDSS J0941 exhibits double-peaked structures consistently across its main narrow emission lines, which are therefore commonly regarded as strong dual AGN candidates \citep{Xu09,Zh24}. As summarized in Table~\ref{tab1}, the peak separations derived from both the SDSS and DESI spectra are mutually consistent among different emission lines, lending further support to this interpretation.
Unfortunately, as shown in Fig. \ref{fig2}, the image reveals no clear signs of dual nuclei or a merger. This can be attributed to inadequate spatial resolution or the blending of the two nuclei. 
Given that the putative AGNs are expected to be separated by kiloparsec-scale distances, their orbital periods are likely on the order of millions of years. As a result, the time baseline of several decades between the SDSS and DESI observations is negligible compared with the characteristic dynamical timescale of the system. 
However, the peak separation measured by SDSS is larger than that by DESI.

In a line-of-sight dual AGN configuration, the Balmer decrements of the blueshifted and redshifted components are 4.48$\pm$0.86 and 3.28$\pm$0.89 in the SDSS spectrum, respectively, and 3.84$\pm$0.68 and 5.26$\pm$2.60 in the DESI spectrum, respectively. These values indicate that neither the foreground nor the background AGN suffers from significant extinction in their NLRs at different aperture scales.
Therefore, a reduction in aperture size is expected to affect both nuclei in a similar manner. However, for SDSS J0941, we found that the line width increases in one component but decreases in the other, and only one component shows a clear decrease in flux. Furthermore, the reduction in aperture size leads to a smaller peak separation is also inconsistent with expectations.
Taken together, these results suggest that SDSS J0941 is unlikely to be explained by a line-of-sight dual AGN configuration.

In the scenario of AGN-driven outflows, given that the aperture centers of all the three spectra are nearly coincident, a reduction in aperture size would be expected to affect the blueshifted and redshifted components in a similar manner. However, the observations of SDSS J0941 do not follow this expectation: the redshifted component shows a clear decrease in both line width and flux, while the blueshifted component instead becomes broader and shows little change in flux.
Dust obscuration within both the host galaxy and the outflowing material itself can preferentially attenuate the receding side of the flow. As a result, the redshifted component is often more heavily extincted and typically exhibits a lower flux compared to the blueshifted component \citep{Liu10,Zh26}. For SDSS J0941, spectral fits to the LAMOST, SDSS, and DESI data consistently show that the redshifted component has a lower flux than the blueshifted component.
In addition, powerful outflows are expected to display radial kinematic stratification, whereby gas at different distances from the nucleus traces distinct velocity regimes \citep{Ba13,Ki20}. 
Under such conditions, different lines within a given spectrum would show different peak separations, and the peak separation of a given line could also vary with aperture size.
However, as shown in Table 1, the peak separations between the blueshifted and redshifted components are relatively consistent across different emission lines within each spectrum, providing no clear evidence of radial kinematic stratification.         
The [O~{\sc iii}] peak separation decreases from $341.4\pm12.0$ km s$^{-1}$ in SDSS to $281.5\pm12.0$ km s$^{-1}$ in DESI, a difference of approximately $3.5\sigma$, whereas the changes for the other lines are less significant given their uncertainties. 
If different gas species exist at different radii, then the flux ratios between different emission lines would also change with aperture size.
We then compared the total flux ratios H$\alpha$/H$\beta$, [N~{\sc ii}]/H$\alpha$, [O~{\sc iii}]/H$\beta$, and ([S~{\sc ii}] $\lambda6716$ plus $\lambda6731$)/H$\alpha$, which differ by only a few percent between the two spectra. The [S~{\sc ii}] $\lambda6716/\lambda6731$ flux ratios likewise show no statistically significant differences between the two velocity components or between the spectra, given the measurement uncertainties. Overall, the available measurements do not allow us to determine whether the observed differences among emission lines and between spectra reflect variations in the spatial distributions or physical conditions of the emitting gas. 
When considering the full set of observational constraints, the asymmetric evolution of the two components is not consistent with the expectations of a simple outflow model.

In the rotating disk-like NLR scenario, the two emission-line components are powered by a single ionizing source, implying that the gas on the approaching and receding sides of the disk should experience comparable physical conditions \citep{Zhang25}. 
However, compared to the SDSS spectrum, one component becomes broader in the DESI spectrum while the other becomes narrower, which is inconsistent with the symmetric expectations of a rotating disk.
Moreover, \citet{Sm12} suggested that a flux ratio between the blueshifted and redshifted components within the range 0.75$\leq$$F_{b}/F_{r}$$\leq$1.25 is characteristic of a rotating disk origin.
To minimize potential contamination from neighboring emission features, we focused primarily on the relatively isolated H$\beta$ and \oiii~emission lines. For SDSS J0941, the flux ratio between the blueshifted and redshifted components derived from the LAMOST spectrum is $1.24 \pm 0.78$ for H$\beta$ and $2.37 \pm 0.19$ for \oiii. The corresponding ratios measured from the SDSS spectrum are $1.40 \pm 0.40$ and $1.99 \pm 0.10$, respectively, while those derived from the DESI spectrum are $3.77 \pm 1.98$ and $3.72 \pm 0.15$. Most of these values deviate substantially from the criterion described above, particularly for \oiii, whose flux ratio is consistently greater than the criterion above. 
Although the smaller aperture yields a reduced peak separation, which appears consistent with naive expectations, the overall observational behavior is not compatible with a simple rotating disk-like NLR model.

\section{Summary and conclusions}
Dual AGN systems, AGN-driven outflows, and rotating disk-like NLR models can all successfully reproduce DPNELs. Therefore, it is difficult to distinguish the physical origin of DPNELs using only a single aperture spectrum.
In this work, we propose a simple and practical method for investigating the origin of DPNELs. The NLR extends from several hundred parsecs to kiloparsecs, and aperture spectra obtained with different aperture sizes probe different spatial regions of the NLR. Since the gas kinematics are expected to vary with spatial scale, the origin of the DPNELs can be constrained by examining the changes in the profiles of the DPNELs observed with different apertures. Moreover, this approach can help mitigate the influence of localized gas kinematic variations on the integrated emission-line profiles of the entire NLR.

Then, we investigate the origin of the DPNELs in SDSS J0941 by comparing spectra obtained with different apertures from the SDSS, DESI, and LAMOST surveys. 
The observed variations in peak separation, line width, and flux between different spectra demonstrate that aperture effects can significantly influence the integrated profiles of DPNELs.
Among the proposed scenarios, the AGN-driven outflow model and rotating disk-like NLR model are disfavored because it cannot naturally account for the strongly asymmetric evolution of the two components.
A dual AGN scenario, in which the two nuclei lie in a plane perpendicular to the line of sight, provides a more self-consistent explanation for the asymmetric evolution.
Nevertheless, the current imaging data provide no direct morphological evidence of dual nuclei or an ongoing merger, and the peak separation measured from the DESI spectrum is slightly smaller than that derived from the SDSS spectrum, which is inconsistent with expectations.

At present, none of the three simple scenarios can satisfactorily reproduce all of the observed properties of the double-peaked profiles. This may indicate that the kinematic structure of the emitting gas is more complicated than that assumed in these idealized models. We therefore explored whether the observed aperture-dependent variations can be understood within more complex scenarios, in particular scenarios in which the blueshifted and redshifted components are not necessarily symmetric and can exhibit different responses to changes in aperture size.
We further considered a more complex rotating NLR with a non-axisymmetric gas distribution. Unlike a simple axisymmetric rotating disk, such a configuration does not necessarily produce symmetric blueshifted and redshifted emission-line components. Instead, differences in the spatial distribution and kinematic properties of the emitting gas can cause the two components to respond differently to changes in aperture size.
If the blueshifted component originates predominantly from a relatively compact region located closer to the center and the redshifted component arises from a more extended region at larger distances, the two components can exhibit distinct aperture-dependent behaviors. As the aperture decreases, the compact blueshifted emitting region remains largely enclosed, resulting in little change in its flux. At the same time, the smaller aperture preferentially samples the inner, higher-velocity gas, which can produce a broader observed line profile. In contrast, the more extended redshifted emitting region is more strongly affected by the reduced aperture. Consequently, a substantial fraction of its emission is excluded, leading to a significant decrease in the observed flux. Furthermore, if the outer part of the redshifted NLR contributes a broader range of projected velocities because of noncircular motions or a more complex velocity field, excluding this extended emission may preferentially suppress the line wings and produce a narrower observed line profile.

Our results illustrate the potential of aperture-dependent spectral variability as a practical diagnostic for investigating the physical origin of DPNELs, particularly for large spectroscopic samples for which high-spatial-resolution observations are not readily available.
At the same time, our test case, SDSS J0941, highlights the complexity of interpreting DPNELs: none of the three proposed scenarios can be unambiguously favored over the others, suggesting that the physical origins of DPNELs are more complex than those described by these basic models.
For SDSS J0941, the observed aperture-dependent variations prefer an elliptical disk-like NLR scenario. Future high-spatial-resolution imaging and integral-field spectroscopic observations would provide a direct test of this interpretation and further evaluating the diagnostic utility of aperture-dependent spectral variability.

\begin{acknowledgements}
We gratefully acknowledge the referee for giving us constructive comments and suggestions to greatly improve our paper.
This work is supported by the National Natural Science Foundation of China (Nos. 12273013, 12173020, 12373014). The SDSS DR16 website is (\url{http://skyserver.sdss.org/dr16});
the DESI website is (\url{https://www.desi.lbl.gov/}); and the LAMOST website is (\url{https://www.lamost.org/}).
The MPFIT website is (\url{http://cow.physics.wisc.edu/~craigm/idl/idl.html}).
\end{acknowledgements}


\begin{thebibliography}{}
\bibitem[\protect\citeauthoryear{Ahumada et al.}{2020}]{Ah20}
Ahumada, R., Allende Prieto, C., Almeida, A., et al., 2020, ApJS, 249, 3
\bibitem[\protect\citeauthoryear{Barrows et al.}{2013}]{Ba13}
Barrows, R. S., Sandberg Lacy, C. H., Kennefick, J., et al, 2013, ApJ, 769, 95
\bibitem[\protect\citeauthoryear{Barth et al.}{2008}]{Ba08}
Barth, A. J., Bentz, M. C., Greene, J. E., \& Ho, L. C. 2008, ApJL, 683, 119
\bibitem[\protect\citeauthoryear{Bellm et al.}{2019}]{Be19}
Bellm, E. C., Kulkarni, S. R., Graham, M. J., et al. 2019, PASP, 131, 018002
\bibitem[\protect\citeauthoryear{Bennert et al.}{2002}]{Be02}
Bennert, N., Falcke, H., Schulz, H., et al., 2002, ApJ, 574, 105
\bibitem[\protect\citeauthoryear{Bentz et al.}{2013}]{Be13}
Bentz, M. C., Denney, K. D., Grier, C. J., et al. 2013, ApJ, 767, 149
\bibitem[\protect\citeauthoryear{Bruzual \& Charlot}{1993}]{Br93}
Bruzual, A. G.; Charlot, S., 1993, ApJ, 405, 538
\bibitem[\protect\citeauthoryear{Bruzual \& Charlot}{2003}]{Br03}
Bruzual, G.; Charlot, S., 2003, MNRAS, 344, 1000        
\bibitem[\protect\citeauthoryear{Cappellari}{2017}]{ca17}
Cappellari, M., 2017, MNRAS, 466, 798
\bibitem[\protect\citeauthoryear{Cheng et al.}{2025}]{Ch25}
Cheng, P. Z., Chen, X. Q., Liao, G. L., et al., 2025, A\&A, 699, 189
\bibitem[\protect\citeauthoryear{Chen et al.}{2025}]{Ch25}
Chen, X., Cheng, P., Gu, Y., et al., 2025, MNRAS, 539, 24
\bibitem[\protect\citeauthoryear{Cid Fernandes et al.}{2005}]{Ci05}
Cid Fernandes, R.; Mateus, A.; Sodré, L, et al., 2005, MNRAS, 358, 363
\bibitem[\protect\citeauthoryear{Comerford et al.}{2013}]{Co13}
Comerford, J. M., Schluns, K., Greene, J. E., \& Cool, R. J. 2013, ApJ, 777, 64
\bibitem[\protect\citeauthoryear{Comerford et al.}{2015}]{Co15}
Comerford, J. M.; Pooley, D.; Barrows, R. S., et al., 2015, ApJ, 806, 219
\bibitem[\protect\citeauthoryear{Comerford et al.}{2018}]{Co18}
Comerford, J.M.; Nevin, R.; Stemo, A., et al, 2018, ApJ, 867, 66
\bibitem[\protect\citeauthoryear{Crenshaw \& Kraemer}{2000}]{Cr00}
Crenshaw, D. M., \&, Kraemer, S. B., 2000, ApJ, 532, 101
\bibitem[\protect\citeauthoryear{Cui et al.}{2012}]{Cu12}
Cui, X. Q., Zhao, Y. H., Chu, Y. Q., et al., 2012, RAA, 12, 1197
\bibitem[\protect\citeauthoryear{Das et al.}{2005}]{Da05}
Das, V., Crenshaw, D.~M., Hutchings, J.~B., et al., 2005, AJ, 130, 945
\bibitem[\protect\citeauthoryear{Das et al.}{2005}]{Da06}
Das, V., Crenshaw, D.~M., Kraemer, S.~B., et al., 2006, AJ, 132, 620 
\bibitem[\protect\citeauthoryear{Deconto-Machado et al.}{2022}]{De22}
Deconto-Machado, A., Riffel, R.~A., Ilha, G.~S., et al., 2022, A\&A, 659, 131
\bibitem[\protect\citeauthoryear{Dey et al.}{2019}]{De19}
Dey, A., Schlegel, D. J., Lang, D., et al. 2019, AJ, 157, 168
\bibitem[\protect\citeauthoryear{Fischer et al.}{2013}]{Fi13}
Fischer, T.~C., Crenshaw, D.~M., Kraemer, S.~B., et al., 2013, ApJS, 209, 1 
\bibitem[\protect\citeauthoryear{Fu et al.}{2012}]{Fu12}
Fu, H.; Yan, L.; Myers, A.D., et al., 2012, ApJ, 745, 67
\bibitem[\protect\citeauthoryear{Ge et al.}{2012}]{Ge12}
Ge, J. Q., Hu, C., Wang, J. M., et al, 2012, ApJS, 201, 31
\bibitem[\protect\citeauthoryear{Hainline et al.}{2014}]{He14}
Hainline, K. N., Hickox, R. C., Greene, J. E., et al., 2014, ApJ, 787, 65
\bibitem[\protect\citeauthoryear{Husemann et al.}{2022}]{Hu22}
Husemann, B., Singha, M., Scharw{\"a}chter, J., et al., 2022, A\&A, 659, 124
\bibitem[\protect\citeauthoryear{Kauffmann et al.}{2003}]{Ka03}
Kauffmann, G.; Heckman, T. M.; Tremonti, C., et al., 2003, MNRAS, 346, 1055
\bibitem[\protect\citeauthoryear{Kewley et al.}{2001}]{Ke01}
Kewley, L. J.; Dopita, M. A.; Sutherland, R. S., et al., 2001, ApJ, 556, 121
\bibitem[\protect\citeauthoryear{Kim et al.}{2020}]{Ki20}
Kim, D. C.; Yoon, I.; Evans, A.~S., 2020, ApJ, 904, 23
\bibitem[\protect\citeauthoryear{King \& Pounds}{2015}]{Ki15}
King, A. \& Pounds, K., 2015, ARA\&A, 53, 115
\bibitem[\protect\citeauthoryear{Lebowitz et al.}{2025}]{Le25}
Lebowitz, S., Hainline, K., Juneau, S., et al., 2025, ApJ, 984, 13
\bibitem[\protect\citeauthoryear{Liu et al.}{2010}]{Liu10}
Liu, X.; Shen, Y.; Strauss, M.A., 2010, ApJ, 708, 427
\bibitem[\protect\citeauthoryear{Liu et al.}{2010b}]{Liu10b}
Liu, X.; Greene, J. E.; Shen, Y., et al., 2010b, ApJL, 715, 30
\bibitem[\protect\citeauthoryear{Liu et al.}{2013}]{Li13}
Liu, G., Zakamska, N. L., Greene, J. E., et al., 2013, MNRAS, 430, 2327
\bibitem[\protect\citeauthoryear{Marconcini et al.}{2025}]{Ma25}
Marconcini, C., Marconi, A., Cresci, G., et al., 2025, NatAs, 9, 907
\bibitem[\protect\citeauthoryear{Markwardt}{2009}]{Ma09}
Markwardt, C.~B., 2009, ASPC, 411, 251
\bibitem[\protect\citeauthoryear{McGurk et al.}{2015}]{Mc15}
McGurk, R.~C.; Max, C.~E.; Medling, A.~M., et al., 2015, ApJ, 811, 14
\bibitem[\protect\citeauthoryear{Noda \& Done}{2018}]{No18}
Noda, H., \& Done, C., 2018, MNRAS, 480, 3898
\bibitem[\protect\citeauthoryear{Panda \& {\'S}niegowska}{2024}]{Pa24}
Panda, S., \& {\'S}niegowska, M., 2024, ApJS, 272, 13
\bibitem[\protect\citeauthoryear{Remigio et al.}{2025}]{Re25}
Remigio, R. P., U, V., Barth, A. J., et al., 2025, ApJ, 992, 42
\bibitem[\protect\citeauthoryear{Peterson et al.}{2013}]{Pe13}
Peterson, B.~M., Denney, K.~D., De Rosa, G., et al., 2013, ApJ, 779, 109
\bibitem[\protect\citeauthoryear{Ricci \& Trakhtenbrot}{2023}]{Ri23}
Ricci, C., \& Trakhtenbrot, B., 2023, NatAs, 7, 1282
\bibitem[\protect\citeauthoryear{Polack et al.}{2024}]{Po24}
Polack, G. E., Revalski, M., Crenshaw, D. M., et al., 2024, ApJ, 975, 129
\bibitem[\protect\citeauthoryear{Ruiz et al.}{2005}]{Ru05}
Ruiz, J.~R., Crenshaw, D.~M., Kraemer, S.~B., et al., 2005, AJ, 129, 73
\bibitem[\protect\citeauthoryear{Runnoe et al.}{2016}]{Ru16}
Runnoe, J. C., Cales, S., Ruan, J. J., et al. 2016, MNRAS, 455, 1691
\bibitem[\protect\citeauthoryear{Rubinur et al.}{2019}]{Ru19}
Rubinur, K.; Das, M.; Kharb, P., 2019, MNRAS, 484, 4933
\bibitem[\protect\citeauthoryear{Severgnini et al.}{2021}]{Se21}
Severgnini, P.; Braito, V.; Cicone, C., et al, 2021, A\&A, 646, 153  
\bibitem[\protect\citeauthoryear{Shen et al.}{2011}]{Sh11}
Shen, Y.; Liu, X.; Greene, J.E., et al, 2011, ApJ, 735, 48
\bibitem[\protect\citeauthoryear{Smith et al.}{2012}]{Sm12}
Smith, K. L., Shields, G.A., Salviander, S., et al, 2012, ApJ, 752, 63
\bibitem[\protect\citeauthoryear{Stern et al.}{2021}]{St18}
Stern, D., McKernan, B., Graham, M. J., et al., 2018, ApJ, 864, 27
\bibitem[\protect\citeauthoryear{Villforth \& Hamann}{2015}]{Vi15}
Villforth, C.; Hamann, F., 2015, AJ, 149, 92
\bibitem[\protect\citeauthoryear{Walsh et al.}{2008}]{Wa08}
Walsh, J. L., Barth, A. J., Ho, L. C., et al., 2008, AJ, 136, 1677
\bibitem[\protect\citeauthoryear{Wang et al.}{2018}]{Wa18}
Wang, J., Xu, D. W., \& Wei, J. Y. 2018, ApJ, 858, 49
\bibitem[\protect\citeauthoryear{Xu \& Komossa}{2009}]{Xu09}
Xu, D. W., Komossa, S., 2009, ApJ, 705, 20
\bibitem[\protect\citeauthoryear{Zhao et al.}{2012}]{Zh12}
Zhao, G., Zhao, Y. H., Chu, Y. Q., et al., 2012, RAA, 12, 723
\bibitem[\protect\citeauthoryear{Zhou et al.}{2004}]{Zh04}
Zhou, H. Y.; Wang, T. G.; Zhang, X. G., et al, 2004, ApJ, 604, 33
\bibitem[\protect\citeauthoryear{Zhang et al.}{2025}]{Zhang25}
Zhang, Z., Chen, Y., Shen, S., et al., 2025, MNRAS, 543, 3089
\bibitem[\protect\citeauthoryear{Zheng et al.}{2023}]{Zh23}
Zheng, Q., Liu, S., Zhang, X. G., et al. 2023, ApJ, 944, 4
\bibitem[\protect\citeauthoryear{Zheng et al.}{2024}]{Zh24}
Zheng, Q., Zhang, X. G., Yuan, Q. R., et al. 2024, MNRAS, 531, L76
\bibitem[\protect\citeauthoryear{Zheng et al.}{2025}]{Zh25}
Zheng, Q., Ma, Y. S., Zhang. X. G., 2025, ApJS, 277, 49 
\bibitem[\protect\citeauthoryear{Zheng et al.}{2026a}]{Zh26a}
Zheng, Q., Zhang, X. G., Yuan, Q. R., 2026a, ApJ, 997, 75
\bibitem[\protect\citeauthoryear{Zheng et al.}{2026b}]{Zh26}
Zheng, Q., Jiang, Y. W., Zhang, X. G., et al., 2026b, A\&A, 710, 31
\end{thebibliography}
\end{document}